\documentclass[12pt]{article}

 \usepackage{amsfonts}
 \usepackage{amssymb,amsmath}
 \usepackage{graphicx} \usepackage{epstopdf}
 \newcommand{\crlb}[1]{\label{#1}\\[2pt]}
 
 \newcommand{\crld}[1]{\label{#1}}
 \newcommand{\eela}[1]{\quad\hbox{\scriptsize{#1}}\label{#1}\end{eqnarray}}
 \newcommand{\eelb}[1]{\label{#1}\end{eqnarray}}
 
 \newcommand{\newsecb}[2]{\section{#1}\label{#2}\setcounter{equation}{0}}

 \newcommand{\nolabels} {\def\eel{\eelb}\def\eeql{\eeqlb}  \def\crl{\crlb} \def\newsecl{\newsecb}\def\bibiteml{\bibitem} \def\citel{\cite}\def\labell{\crld}}
\newcommand{\eeqla}[1]{\quad\hbox{\scriptsize{#1}}\label{#1}\end{aligned}\end{equation}}
\newcommand{\eeqlb}[1]{\label{#1}\end{aligned}\end{equation}}

\newcommand\publishversion{\nolabels\setlength{\textheight}{8.4in}\setlength{\oddsidemargin}{0in}
   	 \setlength{\textwidth}{6.3in}\setlength{\topmargin}{-0.2in}}

\def\beq{\begin{equation}\begin{aligned}}		\def\eeq{\end{aligned}\end{equation}}
\def\be{\begin{eqnarray}}  					\def\ee{\end{eqnarray}}		
\def\bi#1{\begin{itemize}\item[#1]} 			 			\def\ei{\end{itemize}} 
  \def\eqn#1{(\ref{#1})}
   	 \def\fn{\footnote}	  		 

		 \def\a{\alpha}   \def\b{\beta}   \def\d{\delta}          

    		  	\def\D{\Delta}

\def\bra{\langle} 		\def\ket{\rangle}

\def\fract#1#2{{\textstyle\frac{#1}{#2}}}	 	 	
\def\ffract#1#2{\raise .2 em\hbox{$\scriptstyle#1$}\kern-.3em/\kern-.2em\lower .15 em \hbox{$\scriptstyle#2$}}
\def\half{\fract12}

\def\bpmatrix{\begin{pmatrix}} 			\def\epmatrix{\end{pmatrix}}
\def\bmatrix{\begin{matrix}} 			\def\ematrix{\end{matrix}} 
\def\bcenter{\begin{center}}			\def\ecenter{\end{center}}

\def\lowerheightfig#1#2#3{\(\raise-#1\hbox{\includegraphics[height=#2]{#3}}\)}
\def\lowerwidthfig#1#2#3{\(\raise-#1\hbox{\includegraphics[width=#2]{#3}}\)}
				 
\def\weglaten#1{}		
 
  \def\init{{\mathrm{init}}} \def\fin{{\mathrm{final}}}
 \publishversion

\begin{document}

\begin{titlepage}
 \title{{ \LARGE\textbf {The Ontology Conservation Law as an Alternative to the Many World Interpretation of Quantum Mechanics\fn{Presented at the Conference PAFT2019, on `Current Problems in Theoretical Physics',  session on `Foundations of Quantum Theory', Vietri sul Mare, Salerno, Italy, April 13-17, 2019.} \\[20pt]}}}
		\author{Gerard 't~Hooft}
\date{\small  Institute for Theoretical Physics \\ Utrecht University  \\[20pt]
 Postbox 80.089 \\ 3508 TB Utrecht, the Netherlands  \\[20pt]
e-mail:  g.thooft@uu.nl \\ internet: 
http://www.staff.science.uu.nl/\~{}hooft101/  \\[20pt]}
 \maketitle
  \abstract{A sharper formulation is presented for an interpretation of quantum mechanics advocated by author. As an essential element we put forward conservation laws concerning the ontological nature of a variable, and the uncertainties concerning the realisation of states. Quantum mechanics can then be treated as a device that combines statistics with mechanical, deterministic laws, such that uncertainties are passed on from initial states to final states.}
 \end{titlepage}							
  \setcounter{page}{2}
\newsecl{Introduction}{intro}

Deterministic interpretations of quantum mechanics, sometimes called `hidden variable' theories\,\cite{GtH-2010,GtH-2014}, are not strongly supported at present, mainly because the EPR paradox\,\cite{EPR-1935}, Bell's theorem\,\cite{Bell-1964,Bellcuisine}, and related considerations appear to be irreconcilable with  local hidden variables. Although weaknesses in Bell's analysis have been pointed out by many authors\,\cite{Brans-1987}, there appears to be consensus concerning the need to add at least some stochastic element, some `typically quantum' features that cannot be reduced to hidden variables. This means that some version of a `many world' philosophy or pilot wave scenario would be unavoidable; somewhat surprisingly, there appears to be only a minor amount of opposition against such ideas.

We here propose that one should also have a closer look in the opposite direction: if a deterministic theory fails, it might be because it does not take all consequences of deterministic laws into account; it does not carry the notion of determinism far enough. Not only are physical observables constrained by strictly deterministic behaviour, but also observers themselves in their determination to choose what to observe, or more prosaically, their ``free will" is not exactly free, it must also be controlled by deterministic laws. This may imply that `counterfactual' observations may be prohibited by the laws of physics, and considerations concerning the outcomes of counterfactual measurements may have to be handled with much more care than what is usually done, or even be discarded altogether.

A book was written\,\cite{GtH-CA}, in which we bundled our arguments in favour of a completely deterministic approach, and the required techniques are exhibited there. It was made freely available on the internet. We emphasised the elegance of the explanations obtained for Born's probability interpretation of the amplitudes, of the measurement problem, and the Schr\"odinger Cat paradox.

We stress that determinism is only assumed to hold at the tiniest distance scales, typically the Planck length; as soon as one looses these tiny details out of sight, determinism gets lost. 

Nevertheless, our arguments were still not accepted as they appeared to raise yet another problem. When constructing models to investigate the modification needed to our notions of causality in order to exploit this escape route, it seems that some form of `conspiracy' is needed, which again appears to be objectionable. It had been this fear for conspiracy that motivated Bell to consider a definition of causality that is not used in quantised field theories, as it does not hold there. ``Conspiracy is unscientific", is the prevailing opinion. Our book did not take away all such concerns.

Yet the conspiracy problem can also be addressed. We had pointed out that the proposed underlying theories do not exhibit any form of conspiracy at all, but this did not impress the community much. We now think that there is a more powerful way to explain what happens: there is an exact conservation law, the \emph{conservation of ontology}. We assume that the states the universe can be in are divided into two types: the \emph{ontological states} and the \emph{superimposed states}. If the universe is in a superimposed state, this means that we have a probabilistic mixture of different ontologies. It means that it is not specified with certainty which of these ontological descriptions actually apply to the universe; it it due to human ignorance that we do not know exactly which ontological state we are dealing with.

The distribution of these probabilities is absolutely conserved in time. A probabilistic mixture of possible states never evolves into s single state that is realised wit certainty. In short, if the initial state of some process under study is an ontological state, then the final state will be ontological as well; if the initial state is a probabilistic mixture of ontological states, expressed by the superposition coefficients, then so will be the final state, where \emph{the distribution of the superposition coefficients,} and hence also the distribution of the probabilities, is \emph{exactly the same} as those of the initial state. We emphasise that we are referring to superposition coefficients in the ontological basis of Hilbert space.

The conservation of ontology de facto excludes those counterfactual observations that would amount to Alice and/or Bob to `change their minds' as to what settings to choose for their observations, without  also modifying the wave function of the observed entangled photons. In deterministic theories, one cannot modify the state observed at present, without also allowing some changes in the states of the past, regardless whether these describe odd-looking random number generators or some distant quasars whose oscillating photons could be used to choose between Alice's and/or Bob's settings.

\emph{If the results of Alice's and Bob's measurements are ontological, then the quantum state of the observed photons must have been ontological states as well,} from the very beginning, regardless the degree of entanglement these photons may seem to possess.

\newsecl{Causality}{causality}
	Consider an operator \(A(\vec x,t)\), depending on data of a system that are located at point \(\vec x\), and it is acting at time \(t\). Similarly, operator \(B(\vec x\,',t')\) acts at space point \(\vec x\,'\) at time \(t'\). If these two operators do not commute, this implies that \(A\) may bring about an action that affects the way \(B\) acts on the system: \(A\,B\ne B\,A\).
It means that \(A\) emits a signal that can be read by \(B\), \emph{or} \(B\) sends out a signal that can be read by \(A\). This can only happen if \((\vec x\,',t')\)is a space-time point that lies either within the future light cone of \((\vec x,t)\), or within its past light cone, but not if \((\vec x,t)\) and \(\vec x\,',t')\) are space-like separated. Thus, in quantum field theories one generally has
\be \hbox{If }\quad (\vec x-\vec x\,')^2-(t-t')^2>0\quad\ \hbox{then }\quad [A(\vec x,t),\,B(\vec x\,',t')]=0\ . \eel{comm}
We actually call this property `causality'. It has important mathematical consequences for functions such as the `propagator', being the vacuum expectation value \(\bra A(\vec x,t)\,B(\vec x\,',t')\ket\), allowing us to check calculations and theorems in quantum field theory. For the present paper it is important to note that this property stays the same if we flip the arrow of time, \(t-t'\leftrightarrow t'-t\).
This symmetry is due to the fact that the vacuum state, \(|0\ket\), is invariant under time reversal.

To prove his theorem, Bell\,\cite{Bellcuisine} needed a stronger form of causality. Indeed, the definition we employed above holds for quantum systems; for hidden variables, Bell needs a causal relationship that works in future directions differently from the past direction. While the state the observed photons are in, will affect the outcomes of the later measurements, decisions made by Alice and Bob for their choices of settings should not affect these same states. Of course they shouldn't, according to Bell, since Bob and Alice may make their decisions much later than the moment the two photons part from one another, so the photons cannot change at the moment the decisions are made. The kind of determinism needed to have the photons anticipate what kind of polarisers they will encounter in the future, is called `superdeterminism'\fn{The word `superdeterminism' is used in various ways in the literature. Here we use the term to indicate that also observers are subject to deterministic laws of nature. We do \emph{not} opt for `predeterminism', which would mean that some intelligent designer is conspiring with the laws of nature in order to force some desired outcome.}, and this is rejected by Bell and all those who accept his judgment. ``Superdeterminism would be unscientific"; yet it would be unavoidable if the laws of physics at the microscopic scale are to be invariant under time reversal.

It is admitted that there could be some indirect effects. In a fully deterministic world, resembling an automaton, Alice and Bob cannot perform `counterfactual' measurements, and it is true that the universe in which they choose a setting in one particular way, cannot be the same universe in which they have changed their minds. So it \emph{could} be that the state of the photons might be different in these different universes, but the mechanism at work seemed to be totally obscure, and therefore it is often thought that this cannot be part of a serious physical mechanism.

The situation may be different however if an exact \emph{conservation law} would dictate the photons as well as settings chosen by Alice
and Bob to be in ontological states only. This we elucidate in the next chapter.

\newsecl{Two conservation laws}{cons}
	Consider the prototype of a deterministic theory. The picture one would like to draw is a universe allowing for a strictly finite number of states to be in. For simplicity one could think of a universe with boundary conditions in each of the three directions in 3-space, making it completely compact. If many particles may occur in this universe, each being allowed to be in one of a finite but huge number of distinct states, the total number of states for the entire universe will be gigantic, but it will be finite for sure. We could label them \(|1\ket,\,|2\ket,\,|3\ket,\,\dots\)\,. 
	A deterministic theory would posit that at the beat of some clock, called the \emph{time parameter} (also discrete, for simplicity), a \emph{permutation} takes place. The complete description of this permutator would be the task of a physicist; it would be nothing less than the `theory of everything', for instance:
	\be P|1\ket = |2\ket,\quad P|2\ket=|3\ket, \quad \hbox{etc.,}  \quad P|N\ket=|1\ket\ ,\eel{pemutate}
where the period \(N\) may be a very large number.
Mathematically, one could identify the states \(|n\ket\) with the elements of a \emph{basis} of a Hilbert space. In this basis, the permutator \(P\) can be identified as the evolution operator \(U(\d t)\) for the system, a matrix containing only ones and zeros. Here, \(\d t\) is the duration of a single beat of the external clock.

It is easy to diagonalise this matrix. Now the matrix clearly preserves the norm of the basis elements, therefore, its eigen values are complex number of norm one. The matrix preserves the norm of \emph{any} vector in this Hilbert space, if the `law' just formulated is a pure permutator, which means that when two initial states are distinguishable, then also the corresponding final states are distinguishable, or, the evolution law is time-reversible. We usually assume time-reversibility, although in principle, such an assumption can be dropped; this would lead to complications that we shall not go into\,\cite{GtH-CA}. 

It is then easy to define the eigen vectors and eigen values of this matrix, and in the basis of the eigen states it is
 also easy to identify an operator \(H\) such that
	\be U(\d t)=P=e^{-iH\,\d t}\ . \eel{hamilton}
Now note that any unitary transformation to a different basis leaves Eq.~\eqn{hamilton} unchanged. It is the theory's Schr\"odinger equation. Here already, one concludes that all deterministic theories, in particular the time-reversible ones, allow a notation as in quantum mechanics, with Hamiltonian~\eqn{hamilton}.

One may use any orthonormal basis for Hilbert space, but only the original basis elements, \(|1\ket,\ |2\ket,\ \dots\,\), will be called \emph{ontological} states. These are the only states our universe will be in, at any instant of time that is an integer multiple of \(\d t\).

Now we observe that one may also consider states that are superpositions. Let the initial state be of the form
	\be\psi_\init=\a|1\ket+\b|2\ket+\cdots\ ,\quad\hbox{with}\quad |\a|^2+|\b|^2+\cdots =1\ . \eel{superimpose}
Then, since ontological states evolve into ontological states, the final state, after any amount of time \(\D t=N\d t\), will take the form
	\be\psi_\fin=\a|\psi_1\ket+\b|\psi_2\ket+\cdots\ , \eel{finalstate}
where \(|\psi_1\ket,\ |\psi_2\ket\) are again ontological states, while the coefficients \(\a,\ \b,\ \dots\) are \emph{the same coefficients} as those of the initial states, Eq.~\eqn{superimpose}.

One is now free to interpret the coefficients \(\a,\ \b,\ \dots\) as describing the probabilities, \(W_1=|\a|^2,\ W_2=|\b|^2,\ \dots\) associated to the initial states \(|1\ket,\ |2\ket,\cdots\). What we just learned above is that \emph{the final ontological states} \(\psi_1,\ \psi_2,\ \dots\) of Eq.~\eqn{finalstate} come with \emph{the same probabilities.} This simple observation is often overlooked, but is is very important.

It is often claimed that, while the initial state was specified completely (for instance a perfectly collinear beam of particles), the final states will nevertheless come with a probability distribution, which indeed is a primary feature of quantum mechanics. But analysing the above mathematics yields the flaw of such an argument: \emph{The initial state cannot have been an ontological state}. Indeed, if particles form a beam with given momenta, then in reality these are superpositions of ontological states. A particle in a pure momentum eigenstate cannot be assumed to be an ontological state; it is a probabilistic distribution.\fn{Note that, particles in their position states \(|\vec x\ket\) are not ontological either, since the position operators at different times do not commute.}  We elaborate on this feature in Chapter \ref{init}.

The conservation of the superposition variables \(\a,\ \b,\ \dots\) is so important that we classify this as a fundamental conservation law, the \emph{conservation of ontology}. If one of the coefficients is one (with possibly a phase factor added) then orthonormality forces all other coefficients to vanish, so that we are dealing with state \(|1\ket\) with certainty. This same certainty will then characterise the final state,\emph{obtained by applying the Schr\"odinger equation without any modification.} Or: ontological in equals ontological out.

We claim that the conservation law is responsible for what happens in any Bell-like experiment. Take Bell's set-up. Let Alice be using a polariser with orientation \(a\), and Bob uses orientation \(c\). The two photons considered in the experiment are assumed to be perfectly entangled.\fn{One sometimes wonders whether \emph{any} entangled state can alway be produced in an experimental set-up. In practice, this always appears to be the case, so that we exclude `forbidden entanglements' as candidates for a loop hole around Bell.} Consequently, one may assume the polarisation angle \(c\) to be always equal for the two photons. 

One then finds that, if there is a hidden variable theory at all, the angles \(a,\ b,\) and \(c\) must be correlated. A simple model calculation finds the probability distribution \(W(a,\,b,\,c)\) to have, at least, the following structure:
	\be W(a,b,c)=C|\sin(4c-2a-2b)|\ , \eel{mousedrop}
where the factor \(C\) normalises the integrated probabilities to one, which depends on details for the coordinates that we need not go into. We proposed to call the function \(W(a,b,c)\) the `mousedrop function'\,\fn{The name is derived from an objection brought forward by a colleague in a discussion on `conspiracy' (unfortunately I do not have a record on who this was). He suggested that both Alice and Bob carry with them a cage with a mouse inside. They choose their settings \(a\) and \(b\) after counting the mouse droppings they see: if the number is odd they choose one angle, if it is even they pick another angle. In this case, the detected photon must have the means to anticipate how the guts of a mouse will work. Isn't this ridiculous? Our point here is merely: also the guts of a mouse obeys ontology conservation, and this rule is all we need to know.}.

This function tells us that, as soon as Alice's setting \(a\) and Bob's setting \(b\) are fixed, the probability distribution \(W\) of the photon polarisation \(c\) is no longer even. The photon must have, ahead of time, at least some knowledge of the settings \(a\) and \(b\). This was against Bell's verdict that Alice and Bob should both have `free will' to choose their settings \(a\) and \(b\), even if the photons were already on their way to the detectors, without the option to affect one another.

But the ontology conservation law tells us that, if both Alice and Bob made their measurements, each of them are dealing with an ontological state. They now know that the polarisation of their photon is in a state with coefficients \(\a\) and \(\b\) that are either 1 or zero, but not in a superposition. The initial photon must now also have been in an ontological state. If Alice and/or Bob \emph{change} their settings, then the photons \emph{must} enter in a different ontological state as well. This is the mere fact that, if Alice and Bob change their settings, the universe they are in is described entirely in terms of ontological states that differ from what they were before. This is why the photons then \emph{have to} enter the scenery in different ontological states as well.

We observe that the ontology conservation law is closely related to its counterpart, \emph{conservation of uncertainty}: if several of the superposition coefficients for the initial state differ from zero, then also in the final state they differ from zero, since these coefficients keep the same values.

But remember that these conservation laws only hold when applied to ontological states. Applicability of these laws is limited by this. The conservation laws serve primarily to understand what happens to `reality' in experiments such as the EPR Gedanken experiment and numerous other apparent quantum `paradoxes'. Until we succeed in finding realistic models for the ontological states. As was explained at length in Ref.\,\cite{GtH-CA}, ontological models require the identification of `beables', which are here taken to be a complete set of observables that all commute with one another, at all times. An example is the `ontological massless fermion model'.

\newsecl{Initial and final states}{init} 
The ontological conservation law and the uncertainty conservation law may well be applied as pawns in the discussion of `reality' in quantum mechanics. What our research strongly suggests is, 
that the basis elements of Hilbert space normally in use to describe a quantum mechanical process, do not represent `alternative realities' forming a gigantic `multiverse', as the `Many World Interpretation' wants to have it, but instead they represent \emph{possible} worlds in imprecise descriptions. Only one world is real, but physicists today are unable to identify its ontological states precisely. The basis elements normally in use are quantum superpositions of all possibilities. A beam of particles in eigenstates of the momentum operator, for instance, is in fact a gigantic superposition of particles in ontological states, where we have been unable to pinpoint exactly how to describe these states.

The superposition coefficients represent the probabilities that the state we work with is actually one of the given ontological states.  During the entire evolution process, these coefficients are conserved (the uncertainty conservation), so that we find a final state that (again) comes with probability distributions. Nowhere in such processes would there be any need for the wave function to `collapse'. Collapse comes free of charge from the evolution equation -- the Schr\"odinger equation. To guarantee the collapse all that is needed is that the initial state is actually one of the given ontological states.

Each of the ontological states are superpositions of the conventional basis elements of Hilbert space. This means that in theories such as the Standard Model of the sub-atomic particles, we should be able to identify the superpositions that end up behaving as the ontological states. They are characterised as operators that commute with one another at all times.

Why do we have quantum mechanics? We may now come with a tentative answer. In our view, the world at microscopic scales is exactly as `classical' as Newtonian mechanics is, but there is one fundamental complication. Whenever we attempt to `explain' a microscopic process, we have to deal with the difficulty that the \emph{initial state} is unknown. How could we know the initial state? The measurements have not yet been started. When we start to measure things, we will have to use other microscopic processes, typically using photons to register the atoms and fields that we have. However, these photons will disturb the system we are looking at. We also cannot characterise these photons with infinite precision because these too start from  unknown initial states.

Thus quantum mechanics may be the answer to the question how to deal with these fundamental uncertainties. We are able to put the initial particles in beams or localised at predesignated spots, but we cannot choose a completely ontological initial state. As the kinetic laws always include reaction after action, we are forced to add statistical elements to our description of states.

If, however, the system we wish to study comes with so many atoms that the microscopic statistical uncertainties cancel out, then we enter into a situation where the initial state is sufficiently well-known. The reaction due to single photons that are used for observations can then be neglected. This is what the `classical limit'  is about. For example, when we study the orbits of planets in the solar system, we may ignore their reactions to the photons used to observe them. This means that the orbital parameters are classical up to a large number of decimal places. If too much accuracy is required, the effects of quantum superposition will be felt again. In practice, planetary orbits, dead and live cats, and measuring apparatus, can be characterised sufficiently accurately so that classical physics applies.

Apart from the fundamental uncertainty of the initial states, which always continues to blur our vision when following a quantum -- that is, microscopic -- process, one also may have uncertainties due to different causes. Thus, on top of the superposition coefficients of the wave function, one may have `ordinary' probabilities; these cause wave functions to mix further,  generating density matrices such as the ones used in thermodynamics.

It is important to note that the ontological states needed to describe the micro world, are likely to differ in many ways from the classical laws we are used to, in spite of their deceptively simple deterministic behaviour. For instance, a particle with spin \(s=0,\ \half,\ 1,\ \dots\,\), has not only \(2s+1\) angular momentum states, but also its orientation in 3-space can be seen to be constrained by not more than \(2s+1\) possible positions.  \\[10pt]

\noindent {\Large\textbf{Acknowledgment}}\\[10pt]
The author benefited from numerous discussions with colleagues, even if no complete agreement was reached. In particular we owe to  M.~Blasone, H.-T.~Elze, J.~Fr\"olich, T.~Maudlin, T.~Norsen, F.~Scardigli, G.~Vitiello, and C.~Wetterich.

\end{document}